\documentclass[twocolumn,showpacs,amssymb,prd]{revtex4}

\usepackage{graphicx}
\usepackage{dcolumn}
\usepackage{bm}
\usepackage{epsfig}

\begin{document}

\title{Solution To The Cosmic Rays  Puzzle ?}

\author{Shlomo Dado and Arnon Dar}
\affiliation{Physics Department, Technion, Haifa 32000, Israel}

\begin{abstract}
	
Recent observations provide compelling evidence that the bulk of the high 
energy cosmic rays (CRs) and gamma-ray bursts (GRBs) are co-produced by 
highly relativistic jets of plasmoids of stellar matter. These jets are 
launched by fall back matter on newly born neutron stars and stellar black 
holes in core collapse of stripped envelope massive stars with or without 
an associated supernova. The electrons in the plasmoids produce GRB pulses 
mainly by inverse Compton scattering of photons on their path, while 
magnetic reflection of the charged particles produces the high energy 
cosmic rays.

\end{abstract}

\maketitle

Cosmic rays (CRs) are mostly high energy, stable, charged particles 
(protons, nuclei and electrons)  which reside in the interstellar and 
intergalactic space. They were discovered in 1912 by Victor Hess 
[1]. Their scattering by interstellar and intergalactic magnetic fields so 
far has prevented identification of their main sources, and the origin of 
their high energies is still debated. In 1949 Fermi suggested [2] that 
their high energies are acquired by being reflected from interstellar 
"magnetic mirrors" - magnetized clouds, which move slowly in random 
directions in the interstellar medium. However, CR particles may loose 
energy by synchrotron radiation faster than they gain by repeated magnetic 
reflections. Consequently, the original Fermi acceleration mechanisml has 
been replaced by the so called Fermi shock acceleration [3-9]. In 
this model charged particles are assumed to gain energy by being scattered 
repeatedly between the upstream and downstream regions of strong shocks 
produced, e.g., by supernova shells expanding into the interstellar 
medium. This shock acceleration mechanism is widely believed 
to be the main origin of galactic and extragalactic cosmic rays.

An alternative model of CR acceleration [10-15], later called the 
cannonball (CB) model, unified the production of cosmic ray bursts (CRBs) 
and gamma ray bursts (GRBs). In this cannonball model, highly relativistic 
jets of plasmoids (CBs) of ordinary stellar matter are launched by fall 
back matter on a newly born neutron star or a stellar black hole in core 
collapse explosion of stripped envelope massive stars. GRBs are produced 
by inverse Compton scattering (ICS) of light photons on the path of the 
jet by the electrons in the plasmoids [16,17], while magnetic reflection 
of the charged paricles by the plasmoids produce the high energy cosmic 
rays [10-15]. In the CB model, the CR knee is the maximum energy that CR 
particles of a given type (electrons, protons or nuclei) acquire in a 
single magnetic reflection. These knee energies depend only on the largest 
Lorentz factor of the plasmoids in such jets and on the mass of the CR 
particles. In the CB model, CRs with energy above their knee are CRs which 
were reflected backward from slower CBs or supernova shells which were 
ejected earlier. This interpretation is different from that adopted in the 
Fermi/shock acceleration models, where the CR knee depends on their 
rigidity $R=pc/Z$, namely on the momentum of the CR particle multiplied 
by the speed of light per unit charge.

The energy spectrum of high energy CRs from well below to well above the 
CR knee is shown in Figure 1 adopted from [18].  
\begin{figure}[]
\centering 
\epsfig{file=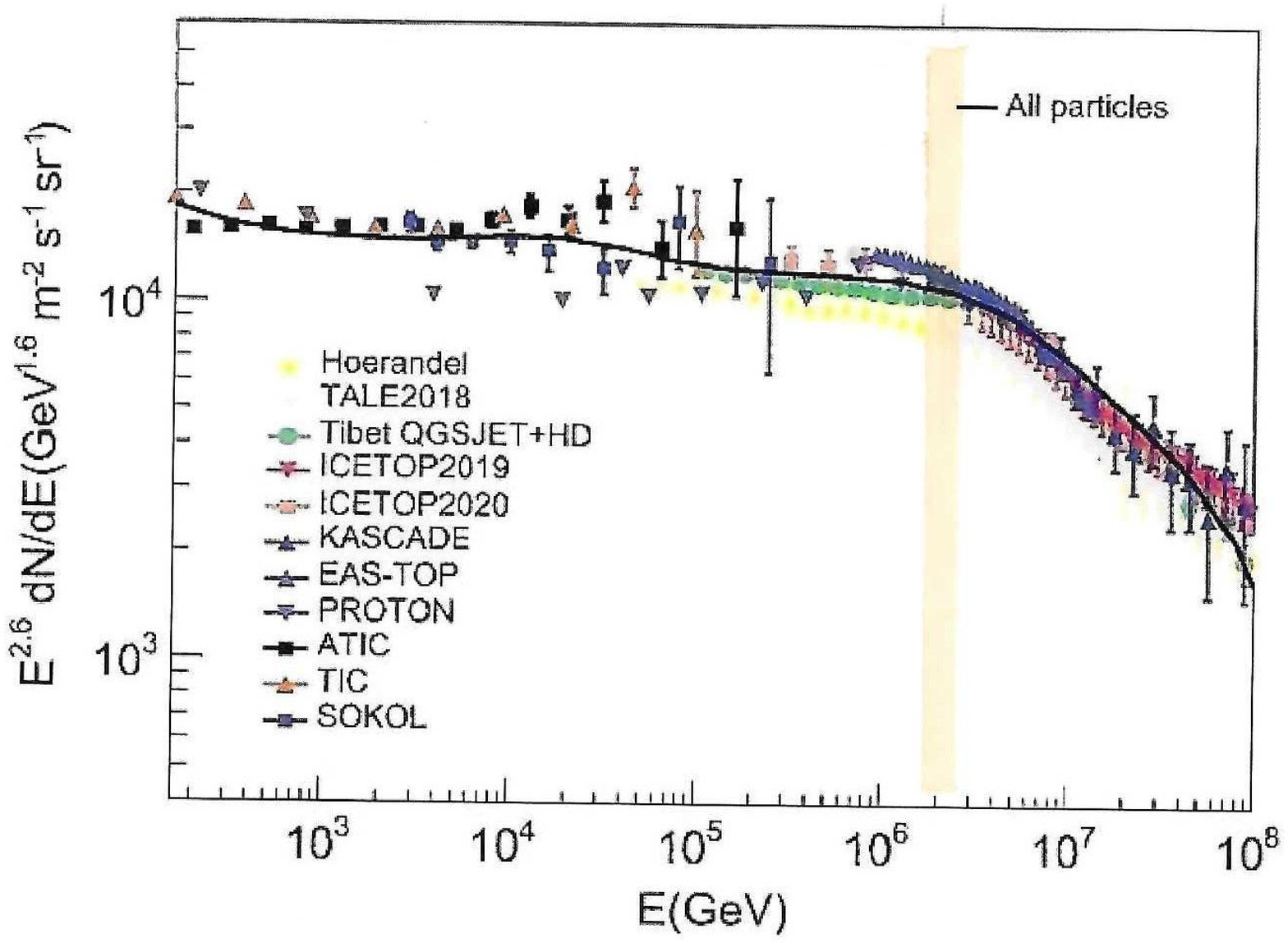,width=8.6cm,height=8.6cm}
\caption{The energy spectrum of cosmic ray nuclei around the cosmic ray
knee reported in [18]. The knee energy of cosmic ray protons is indicated 
by the wide band around 2 PeV}. 
\end{figure}\\

Until recently, the measured knee energies of individual cosmic ray nuclei 
were not accurate enough to conclude whether they depend on their masses, 
as expected in the CB model [13], or on their rigidities as expected 
in the Fermi/shock acceleration models. However, while the rigidities of 
high energy electrons and protons are practically equal, their masses are 
very different; $m_p/me\approx 1836$. In the CB model, that implied  
knee energies of high energy CR electrons which satisfy
[13-15],
\begin{equation}
E_{knee}(e)\approx (m_e/m_p)E_{knee}(p)\approx 1~{\rm TeV}. 
\end{equation}

Fortunately, during the past decade, precise enough measurements of the 
energy spectrum of CR electrons were extended into the TeV range, in 
particular by the H.E.S.S [19,20], AMS [21], Fermi-LAT [22], DAMPE [23] 
and CALET [24] collaborations. As shown in Figure 2, they have confirmed 
the existence of a knee around $\sim 1$ TeV in the energy spectrum of high 
energy cosmic ray electrons, which was predicted by the CB model [13-15] 
from the observed knee around 2 PeV [18] in the energy spectrum of cosmic 
ray protons.

\begin{figure}[]
\centering 
\epsfig{file=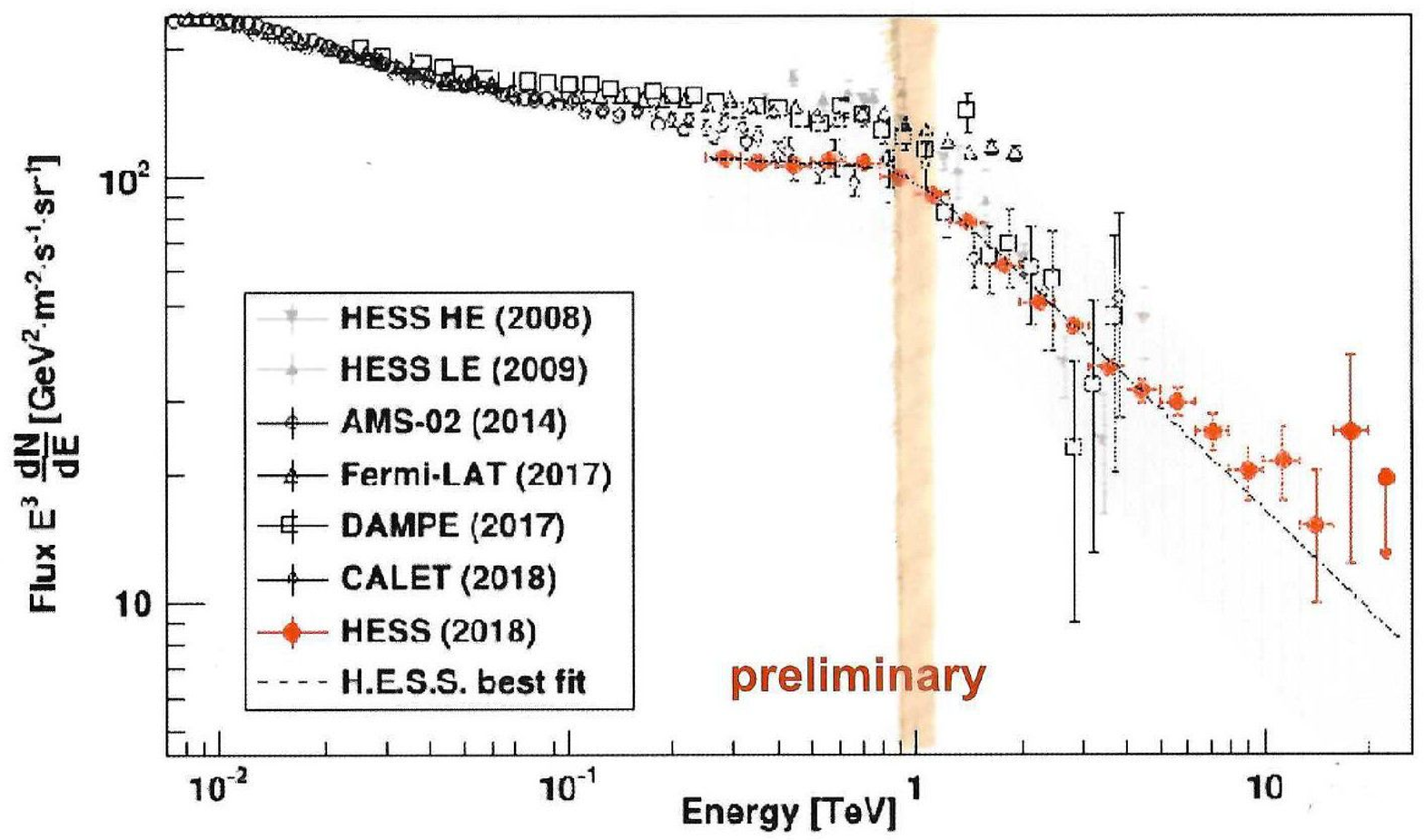,width=8.6cm,height=8.6 cm}
\caption{The high energy spectrum of cosmic ray electrons compiled  in [19]. 
The electron knee energy predicted by the CB model is indicated by the 
vertical band around 1 TeV.}  
\end{figure}

Moreover, the observed knees in the energy spectra of cosmic ray 
nuclei [18] and electrons [19-24] imply that the largest Lorentz 
factor of CBs fired ($t=0$) by the main source of high energy CRs, 
is roughly,
\begin{equation}
\gamma_{max}(0)\approx \sqrt{E_{knee}(CR)/2m_{CR}c^2}\approx 1000.
\end{equation}
In the CB model, this value of $\gamma_{max}(0)$ of CBs at launch is 
common to CR electrons and protons nearly at rest in the CBs. It
alllows two additional critical tests of the joint origin of 
CRBs and GRBs: 

In the CB model, the largest time-averaged peak photon energy of a GRB 
at redshift $z$, which is produced by inverse Compton scattering (ICS)
of optical photons ($\epsilon\approx 1.65$ eV, i.e., 
$\nu\!=\!4\times 10^{14}$ Hz) 
by CB electrons having $\gamma_{max}\approx 1000$, is given by    
\begin{equation}
max [(1+z)E_p]\approx 2(\gamma_{max})^2\epsilon\approx 3.3\,\rm{MeV}.
\end{equation}
This value is consistent with the measured $(1\!+\!z)E_p\!=\!3503\pm133$
keV,[25] at the peak luminosity of the "brightes of all time" GRB 
221009A at redshift $z\!=\!0.151$.

Moreover, the time averaged peak photon energy $E_p\!\approx\!2.912$ MeV 
and the isotropic equivalent energy release, $E_{iso}\approx(1.2\!\pm\!
0.1)\!\times\!10^{55}$ erg measured in GRB 221009A [25] are the record 
high values measured so far in a GRB. Such high values are estimated to 
be observed once in 10,000 years. They were shown [25] to be consistent 
with the best fit Amati correlation [26],
\begin{equation}
(1+z)E_p\propto [E_{iso}]^{0.42},     
\end{equation}
in a sample of 315 Konus-Wind GRBs, which is shown in Figure 3.  
\begin{figure}[]
\centering 
\epsfig{file=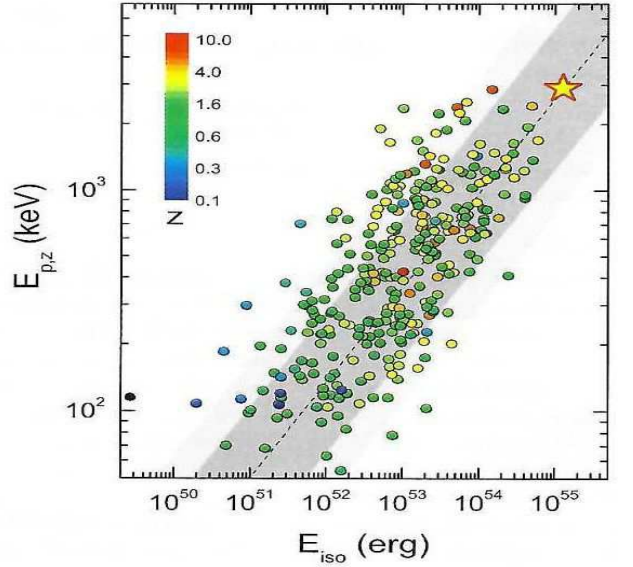,width=8.6cm,height=8.6cm}
\caption{The best fit Amati correlation reported in [25] for
315 long GRBs with known redshift   
observed by Konus-Wind. GRBs are represented by 
circlrs; the color of each data point represents the GRB redshift. 
The error bars are not shown for reasons of clarity.
GRB221009A is indicated by a red star.
The best fit Amati relation is plotted as a dashed line. 
The error bars are not shown  for reasons of clarity.}
\end{figure}

In the CB model [17 for a review], far off axis GRBs, i.e., those 
which are viewd from angles that satisfy, 
$\theta^2\gamma^2\!\gg\! 1$, 
have relatively low $(1\!+\!z)E_p$ and $E_{iso}$ values  which  satisfy,
\begin{equation}
(1+z)E_p\propto [E_{iso}]^{1/3}.
\end{equation}

Near axis GRBs, i.e., those with viewing angles that 
satisfy, $\theta^2\gamma^2 \leq 1$, have relatively large $(1\!+\!z)E_p$ 
and $E_{iso}$ values and satisfy the correlation [26],
\begin{equation} 
(1+z)E_p\propto [E_{iso}]^{1/2}.   
\end{equation}

Consequently, a mixed population of near axis and far off axis GRBs
is expected to satisfy the Amati correlation [26] with an average  
powe-law index $(1/2+1/3)/2\approx 0.42$. Indeed it is that 
reported in [25], and is shown in Figure 3. Moreover, a sum of two 
power laws corresponding to low and high values of $(1\!+\!z)E_p$,     
\begin{equation}
(1+z)E_p=aE_{iso}^{1/3}+ bE_{iso}^{1/2}
\end{equation}
also describes well the mixed population of far off axis GRBs  and 
near axis GRBs.

If the narrow jet of CR protons and nuclei encounters SN shell which was 
ejected earlier, it produces a narrow conical beam of short lived high 
energy pions and kaons along the axis of the much wider GRB cone. Such 
shells have sufficiently low densities for their decay into a narrow 
conical burst of high energy electron and muon neutrinos, electrons and 
gamma rays. Since the tranverse momentum of their $\pi$ and $K$ mesons is 
of the order of their masses [27], their produced high energy neutrinos 
and gamma rays (in the source rest frame) are mainly within a cone of an 
opening angle $\approx m_{\pi}/\gamma m_p$. The high energy gamma rays 
from GRBs are attenuated by pair production on background photons [28], 
while the high energy neutrinos are not. But, both are emitted into a cone 
much narrower than that of the MeV gamma rays from a GRB. That, and the CB 
model estimate [13] of the flux of GRB neutrinos, imply that the chances 
to detect on Earth an high energy (TeV) neutrino burst 
in coincidence with  GRB  MeV photons, are very small,
as was found by IceCube [29]

{\bf Conclusions:} The observed knee energies of cosmic ray protons and 
electrons together with the observed peak energy [25] in the time 
integrated gamma ray spectrum of the "brightest of all time" [30] GRB 
221009A, provide compelling evidence that the bulk of the high energy 
cosmic rays and gamma-ray bursts (GRBs) are co-produced by highly 
relativistic jets of plasmoids of ordinary stellar matter. Such jets are 
launched by fall back matter on newly born neutron stars and stellar black 
holes in core collapse of stripped envelope massive stars, with and 
without an associated supernova. The electrons in the plasmoids produce 
the GRB pulses mainly by inverse Compton scattering of photons on their 
path [16,17], while magnetic reflection of the charged particles produces 
the bulk of the high energy CRs [11-15]. Complete understanding of how 
such highly relativistic jets of plasmoids are formed and why the maximum 
bulk motion Lorentz factor of their plasmoids is $\approx$1000 are still 
lacking.

Acknowledgement: We thank M. Moshe for a useful comment.

{}
\end{document}